# Clusters' size-degree distribution for bond percolation

P. N. Timonin

Southern Federal University, 344091, Rostov-on-Don, Russia

e-mail: pntim@live.ru

*To address some physical properties of percolating systems it can be useful to know the degree distributions in finite clusters along with their size distribution. Here we show that to achieve this aim for classical bond percolation one can use the $q \to 1$ limit of suitably modified q-state Potts model. We consider a version of such model with the additional complex variables and show that its partition function gives the bond percolation's generating function for the size and degree distribution in the $q \to 1$ limit. We derive this distribution analytically for bond percolation on Bethe lattices and complete graph. The possibility to expand the applications of present method to other clusters' characteristics and to models of correlated percolation is discussed.*

## I. INTRODUCTION

Modern percolation theory knows a lot about the structure and geometry of percolation clusters on a number of graphs [1-3]. It is also known how its structure changes with the growth of occupation probability - it acquires high-density [4, 5], and bootstrap (*k*-core) backbones [6, 7] having different sites' degrees (the numbers of bonds attached to a site). In what concerns the finite clusters in percolation, Monte Carlo studies of its average radius, perimeter, fractal dimension, average shape and density profile have been fulfilled for some lattices [8 -12]. Yet most important and most studied is the clusters' size distribution $v_s$ as finding it we get all set of critical indices for percolation transition as well as its order parameter. In its turn, the size distribution $v_s$ is defined by the graph's numbers (per site) $v_{s,t}$ of distinct clusters (lattice animals) with a given size *s* and properly defined perimeter *t* [3, 10, 13]. Thus $v_{s,t}$ (or $v_s$) for a graph is all that needed for standard description of the classical percolation transition on it. So many existing studies of finite clusters in percolation are devoted to numerical determination of $v_{s,t}$ and $v_s$ on various lattices, see, for example, [10, 13-16]. Also $v_s$ is obtained analytically for Bethe lattice [17] and complete graph [18].

The relation of the $q \to 1$ limit of *q*-state Potts model to the $v_s$ generating function [18] is of great help in these studies. Later the modified Potts models were introduced which are related to the generating functions of $v_{s,b}$ [19] and $v_{s,b,t}$ [20] *b* being the number of bonds in a cluster. These works demonstrated for the first time that detailed structural characteristics of finite clusters could be found with the methods of usual statistical mechanics.

To expand this approach one may think of obtaining distributions of other structural characteristics of *s*-site clusters in the Potts model framework. In particular, there are many properties of random systems, which crucially depend on the compactness of finite clusters. They are the ability to participate in chemical reactions, resilience under random removal of their bonds [21] and possibility to acquire net magnetic moment [4, 5], to name a few. To address such problems it is useful to know the degree distributions in finite clusters along with their size distribution. Here we show that to achieve this aim for the classical bond percolation one can use the $q \to 1$ limit of suitably modified *q*-state Potts model. We consider a version of such model with the additional complex variables and show that its partition function gives the bond percolation's generating function for size and degree distribution in the $q \to 1$ limit (Section II). We derive this distribution analytically for bond percolation on Bethe lattices (Section III) and complete graph (Section IV). In Section V we discuss the possibility to expand the applications of present method to other clusters' characteristics and to models of correlated percolation.

## II. MODIFIED POTTS MODEL

Let us first recall the established procedure to find generating function for clusters' size distribution in bond percolation with the $q \to 1$ limit of $q$-state Potts model [18]. For some graph with $N$ sites and $E$ edges on which bonds are placed with probability $p$ consider the Hamiltonian

$$H_0(h,\boldsymbol{\sigma}) = \sum_{\langle i,j \rangle} \ln(1-p)\left(\delta_{\sigma_i,\sigma_j} - 1\right) + \sum_{i=1}^{N} h\left(\delta_{\sigma_i,1} - 1\right)$$

where paired Potts interactions are assigned to each edge of the graph. The partition function of this Potts model is

$$Z_0(q,h) = Tr\, e^{-H_0(h,\boldsymbol{\sigma})} = Tr \prod_{\langle i,j \rangle} \left(1 - p + p\delta_{\sigma_i,\sigma_j}\right) \prod_i e^{h(1-\delta_{\sigma_i,1})}$$

where $Tr$ denotes the sum over all spins $\sigma_i = \{1,...,q\}$. It can be represented as sum over all bond configurations on the graph

$$Z_0(q,h) = \sum_C p^B (1-p)^{E-B} \prod_{clusters} \sum_{\sigma=1}^{q} e^{s_{cl}(1-\delta_{\sigma,1})} = \sum_C p^B (1-p)^{E-B} \prod_s \left[1 + (q-1)e^{sh}\right]^{N_s}$$

Here $B$ is a number of bonds in a given configurations, $s_{cl}$ is a cluster's size and $N_s$ is the number of $s$-site clusters in a configuration. Hence,

$$G_0(h) = \lim_{q \to 1} N^{-1} \frac{d \ln Z_0}{dq} = \sum_s v_s e^{hs}, \quad v_s = \left\langle \frac{N_s}{N} \right\rangle_C$$

Here $\langle ... \rangle_C$ means the average over the random configurations of bonds, which occupy the graph's edges with probability $p$, so $G_0(h)$ is the generating function for the average clusters' size distribution in classical bond percolation.

This derivation relies heavily on the useful property of the model to make equal all the spins in the cluster. Here we intend to use it to obtain the generation function for the degree distribution

$$v_{s_0,\mathbf{s}} = \lim_{N \to \infty} \left\langle \frac{N_{s_0,\mathbf{s}}}{N} \right\rangle_C \quad (1)$$

where $s_0$ and $\mathbf{s} = \{s_1,...,s_z\}$ with $s_k$ defining the number of sites of degree $k$ (i.e., with $k$ bonds attached) in a cluster, $z$ is the maximal number of edges attached to the site in the graph and $N_{s_0,\mathbf{s}}$ is the number of clusters with the given degree distribution $\{s_0,\mathbf{s}\}$ in a given bond configuration. We separate $s_0$ for future convenience as it refers to the special clusters composed of single isolated sites. Apparently, $N_{s_0,\mathbf{s}}$ have the form

$$N_{s_0,\mathbf{s}} = N_1 \delta(s_0, 1) + N_s \delta(s_0, 0)$$

where $N_1$ is the number of single-site clusters and $N_s$ is that of clusters with more than one site which do not have sites with zero degree.

Accordingly, the generation function depend on (*z*+1)-component vector $\{h_0, \mathbf{h}\}$ as follows

$$G(h_0, \mathbf{h}) = \sum_{s_0,\mathbf{s}} v_{s_0,\mathbf{s}} e^{s_0 h_0 + \mathbf{sh}} = v_1 e^{h_0} + \sum_{\mathbf{s}} v_{\mathbf{s}} e^{\mathbf{sh}} = v_1 e^{h_0} + \tilde{G}(\mathbf{h}) \qquad (2)$$

$$v_1 = \lim_{N\to\infty} \left\langle \frac{N_1}{N} \right\rangle_C, \quad v_{\mathbf{s}} = \lim_{N\to\infty} \left\langle \frac{N_{\mathbf{s}}}{N} \right\rangle_C$$

On the regular graphs with single coordination number *z* for all sites

$$v_1 = (1-p)^z$$

so it is sufficient to find $\tilde{G}(\mathbf{h})$ otherwise the full generating function $G(h_0, \mathbf{h})$ should be found. To obtain it we modify original Potts model making it to provide for each cluster the factors $e^{s_k h_k}$. To this aim we introduce the complex variables $\varsigma_i$ to each site and consider the modified Potts Hamiltonian

$$H(\boldsymbol{\sigma}, \boldsymbol{\varsigma}, h_0, \mathbf{h}) = -\sum_{\langle i,j \rangle} \ln\left(1 - p + p \delta_{\sigma_i, \sigma_j} \varsigma_i \varsigma_j\right) - \sum_i \ln f_{\sigma_i}(\varsigma_i, h_0, \mathbf{h}) \qquad (3)$$

and its partition function

$$Z(q, h_0, \mathbf{h}) = Tr e^{-H(\boldsymbol{\sigma}, \boldsymbol{\varsigma}, h_0, \mathbf{h})} = Tr \prod_{\langle i,j \rangle} \left(1 - p + p \delta_{\sigma_i, \sigma_j} \varsigma_i \varsigma_j\right) \prod_i f_{\sigma_i}(\varsigma_i, h_0, \mathbf{h}) \qquad (4)$$

Here

$$Tr A(\boldsymbol{\sigma}, \boldsymbol{\varsigma}) \equiv \sum_{\sigma_1 \ldots \sigma_N} \oint \ldots \oint \frac{d^N \varsigma}{(2\pi i)^N} A(\boldsymbol{\sigma}, \boldsymbol{\varsigma})$$

The integration contours in complex $\varsigma_i$ planes will be specified below.

In bond configuration representation of this model (4) $k_i$ bonds attached to the *i*- th site of a cluster endow it with the factor $\varsigma_i^{k_i}$, $k_i$ being the site's degree. Let us choose the function $f_\sigma(\varsigma, h_0, \mathbf{h})$ with the property

$$\oint \frac{d\varsigma}{2\pi i} f_\sigma(\varsigma, h_0, \mathbf{h}) \varsigma^k = e^{h_k(1 - \delta_{\sigma,1})}, \quad k = \{0, 1, 2, \ldots, z\}, \qquad (5)$$

Then

$$Z(q,h_0,\mathbf{h}) = \sum_C p^B (1-p)^{E-B} \prod_{clusters} \sum_{\sigma=1}^{q} \prod_{i\in cluster} \oint \frac{d\varsigma_i}{2\pi i} f_\sigma(\varsigma_i, h_0, \mathbf{h}) \varsigma_i^{k_i}$$

$$= \sum_C p^B (1-p)^{E-B} \prod_{clusters} \sum_{\sigma=1}^{q} \exp\left[(1-\delta_{\sigma,1}) \sum_{i\in cluster} h_{k_i}\right]$$

If cluster has $s_k$ sites of degree $k$ then

$$\sum_{i\in cluster} h_{k_i} = \sum_{k=0}^{z} s_k h_k$$

and

$$Z(q,h_0,\mathbf{h}) = \sum_C p^B (1-p)^{E-B} \prod_{s_0,\ldots,s_z} \left[1+(q-1)\exp\sum_{k=0}^{z} s_k h_k\right]^{N_{s_0,\ldots,s_z}}$$

Here $N_{s_0,\ldots,s_z} \equiv N_{s_0,\mathbf{s}}$ is the number of clusters with $s_k$ sites of degree $k$, $k=0,\ldots,z$, in the bond configuration. Hence, the model gives in the $q \to 1$ limit

$$G(h_0,\mathbf{h}) = \lim_{q\to 1} \lim_{N\to\infty} N^{-1} \frac{d\ln Z(q,h_0,\mathbf{h})}{dq} = \lim_{N\to\infty} \sum_{\mathbf{s}} \left\langle \frac{N_{s_0,\mathbf{s}}}{N} \right\rangle_C e^{h_0 s_0 + \mathbf{h}\mathbf{s}}$$

Thus, $G(h_0,\mathbf{h})$ in this equation is indeed the generating function for the clusters' degree distributions $\nu_{s_0,\mathbf{s}}$ defined in Eq. (1).

We can choose the precise form of $f_\sigma(\varsigma,h_0,\mathbf{h})$ using the identity

$$\oint_{|\varsigma|>1} \frac{d\varsigma}{2\pi i} \frac{\varsigma^m}{\varsigma-1} = \vartheta(m), \quad \vartheta(0)=1$$

$\vartheta(x)$ is the Heaviside's step function. Thus, $f_\sigma(\varsigma,h_0,\mathbf{h})$ obeying Eq. (5) is

$$f_\sigma(\varsigma,h_0,\mathbf{h}) = \frac{1}{\varsigma-1}\left\{\delta_{\sigma,1} + (1-\delta_{\sigma,1})\left[\sum_{m=0}^{z-1} e^{h_m}\left(\varsigma^{-m} - \varsigma^{-m-1}\right) + e^{h_z}\varsigma^{-z}\right]\right\}$$

$$= \frac{1}{\varsigma-1}\left\{\delta_{\sigma,1} + (1-\delta_{\sigma,1})\left[\sum_{m=1}^{z} \varsigma^{-m}\left(e^{h_m} - e^{h_{m-1}}\right) + e^{h_0}\right]\right\}$$

and integration of it is performed over the circle with radii greater than 1.

The Hamiltonian (3) with this $f_\sigma(\varsigma,h_0,\mathbf{h})$ takes the form

$$H(\boldsymbol{\sigma},\varsigma,h_0,\mathbf{h}) = -\sum_{\langle i,j\rangle}\left[\delta_{\sigma_i,\sigma_j}\ln\left(1+\frac{p\varsigma_i\varsigma_j}{1-p}\right)+\ln(1-p)\right]+$$

$$+\sum_i(\delta_{\sigma_i,1}-1)\ln\left[\sum_{m=0}^{z-1}e^{h_m}\left(\varsigma_i^{-m}-\varsigma_i^{-m-1}\right)+e^{h_z}\varsigma_i^{-z}\right]+\sum_i\ln(\varsigma_i-1) \quad (6)$$

Thus, we have Potts model with the inhomogeneous coupling and field, which depend on the auxiliary complex variables $\varsigma_i$, and with $z+1$ parameters $h_k$. The relation of this statistical mechanics' model to the bond percolation problem can made its study easier in view of the existing arsenal of exact and approximate methods for finding partition functions.

### III. PERCOLATION ON BETHE LATTICE

The advantages of finding the generating function for the degree distribution in bond percolation via partition function of modified Potts model can be fully demonstrated for the graphs allowing its analytical calculation. We consider the Bethe lattice with the coordination number $z$ ($z$ edges are attached to each site). Introducing for our model the partial partition function of the $m$-level tree $Z_\sigma^{(m)}(\varsigma)$ in which trace is made over all sites' variables except the root ones, we get the recurrence relations [22]

$$Z_\sigma^{(m+1)}(\varsigma) = \oint\frac{d\varsigma'}{2\pi i}\sum_{\sigma'=1}^q(1-p+p\varsigma\varsigma'\delta_{\sigma,\sigma'})f_{\sigma'}(\varsigma',h_0,\mathbf{h})\left[Z_{\sigma'}^{(m)}(\varsigma')\right]^{z-1}$$

This equation implies the following form of $Z_\sigma^{(m)}(\varsigma)$

$$Z_\sigma^{(m)}(\varsigma) = a_m + \delta_{\sigma,1}\varsigma(b_m-a_m)+(1-\delta_{\sigma,1})\varsigma a_m u_m \quad (7)$$

so we get the recurrence relations for coefficients in Eq. (7)

$$a_{m+1} = (1-p)\left[b^{z-1}+(q-1)a_m^{z-1}\sum_{k=0}^{z-1}\binom{z-1}{k}u_m^k e^{h_k}\right] \quad (8)$$

$$b_{m+1} = pb_m^{z-1}+a_{m+1} = b_m^{z-1}+(q-1)(1-p)a_m^{z-1}\sum_{k=0}^{z-1}\binom{z-1}{k}u_m^k e^{h_k} \quad (9)$$

$$a_{m+1}u_{m+1} = pa_m^{z-1}\sum_{k=0}^{z-1}\binom{z-1}{k}u_m^k e^{h_{k+1}} \quad (10)$$

The density of the logarithm of full partition function is [22]

$$2\lim_{N\to\infty}N^{-1}\ln Z(q,h_0,\mathbf{h}) = z\lim_{m\to\infty}\left\{\ln Trf_\sigma(\varsigma,h_0,\mathbf{h})Z_\sigma^{(m+1)}(\varsigma)\left[Z_\sigma^{(m)}(\varsigma)\right]^{z-1}\right\}$$

$$-2(z-1)\lim_{m\to\infty}\left\{\ln Trf_\sigma(\varsigma,h_0,\mathbf{h})\left[Z_\sigma^{(m)}(\varsigma)\right]^z\right\} = (2-z)\ln Trf_\sigma(\varsigma,h_0,\mathbf{h})\left[Z_\sigma^{(\infty)}(\varsigma)\right]^z$$

$$= (2-z)\ln\left[b^z+(q-1)a^z\sum_{k=0}^z\binom{z}{k}u^k e^{h_k}\right]$$

Here *a*, *b*, *u* are the stationary points of the recurrence relations (8-10). For $q \to 1$, we have from (8-10)

$$a = 1 - p + O(q-1) \ , \ b \approx 1 + (q-1)\frac{(1-p)^z}{2-z}\sum_{k=0}^{z-1}\binom{z-1}{k}u^k e^{h_k} ,$$

with $u = u(\mathbf{h})$ being the solution to the equation

$$u = p(1-p)^{z-2} A(u,\mathbf{h}) \ , \quad A(u,\mathbf{h}) = \sum_{k=0}^{z-1}\binom{z-1}{k}u^k e^{h_{k+1}} \tag{11}$$

Thus the generating function $G(h_0,\mathbf{h})$ is

$$G(h_0,\mathbf{h}) = G[u(\mathbf{h}),h_0,\mathbf{h}] = \lim_{q\to 1}\lim_{N\to\infty} N^{-1}\frac{d\ln Z(q,h_0,\mathbf{h})}{dq} =$$

$$= \frac{(1-p)^z}{2}\sum_{k=0}^{z}\left[z\binom{z-1}{k} - (z-2)\binom{z}{k}\right]u^k e^{h_k}$$

Using the relations for binomial coefficients we get $G(u,h_0,\mathbf{h})$ in a simpler form

$$G(u,h_0,\mathbf{h}) = \tilde{G}(u,\mathbf{h}) + (1-p)^z e^{h_0} \tag{12}$$

$$\tilde{G}(u,\mathbf{h}) = \frac{z}{2}(1-p)^z \sum_{k=1}^{z}\left(\frac{2}{k}-1\right)\binom{z-1}{k-1}u^k e^{h_k} = \frac{z}{2}(1-p)^z\left[2\int_0^u dv A(v,\mathbf{h}) - u A(u,\mathbf{h})\right] \tag{13}$$

In principle, Eqs. (11-13) solve the problem of finding the averages over clusters' degree distribution as

$$\prod_{k=1}^{z}\frac{d^{m_k}}{dh_k^{m_k}}\tilde{G}[u(\mathbf{h}),\mathbf{h}]\bigg|_{\mathbf{h}=0} = \sum_{\mathbf{s}} \nu_{\mathbf{s}} \prod_{k=1}^{z} s_k^{m_k} \equiv \left\langle \prod_{k=1}^{z} s_k^{m_k} \right\rangle \tag{14}$$

We get all these averages expressed via $u_0 = u(\mathbf{h}=0)$ which obeys the equation

$$u_0 = p(1-p)^{z-2}(1+u_0)^{z-1} \tag{15}$$

For example, mean cluster number per site is

$$\left\langle \frac{N_{cl}}{N}\right\rangle_C = G(u_0,0,0) = \tilde{G}(u_0,0) + (1-p)^z =$$

$$= (1-p)^z\left[1 + z\int_0^{u_0} dv A(v,0) - \frac{z}{2}u_0 A(u_0,0)\right] =$$

$$= (1-p)^z(1+u_0)^{z-1}\left(1 - \frac{z-2}{2}u_0\right) = \frac{(1-p)^2}{p}u_0\left(1 - \frac{z-2}{2}u_0\right)$$

We can express $u_0$ via $S$ - the fraction of sites belonging to the percolation cluster. To do this we should find all $\langle s_k \rangle$. Then summing them we obtain the fraction of sites belonging to all finite clusters, hence

$$\sum_{k=0}^{z}\langle s_k \rangle = 1 - S \qquad (16)$$

We have

$$\frac{d\tilde{G}(u(\mathbf{h}),\mathbf{h})}{d\mathbf{h}} = \left[\partial_u \tilde{G}(u,\mathbf{h})\frac{du(\mathbf{h})}{d\mathbf{h}} + \partial_\mathbf{h} \tilde{G}(u,\mathbf{h})\right]_{u=u(\mathbf{h})} = z(1-p)^z \left\{ \int_0^{u(\mathbf{h})} dv \partial_\mathbf{h} A(v,\mathbf{h}) \right\}$$

Here $\partial_u$ and $\partial_\mathbf{h}$ denote partial derivatives with respect to $u$ and $\mathbf{h}$ while

$$\frac{du(\mathbf{h})}{d\mathbf{h}} = \frac{A(u,\mathbf{h})}{A(u,\mathbf{h}) - u\partial_u A(u,\mathbf{h})}\bigg|_{u=u(\mathbf{h})}$$

So, according to Eq. (14),

$$\langle s_k \rangle = (1-p)^z \binom{z}{k} u_0^k$$

This equation is also valid for $k = 0$, hence

$$\sum_{k=0}^{z}\langle s_k \rangle = \left[(1-p)(1+u_0)\right]^z = 1 - S$$

and $u_0 = \dfrac{p}{1-p}(1-S)^{\frac{z-1}{z}}$, so Eq. (15) turns into the standard Bethe lattice equation for $S$ [17]

$$p(1-S)^{\frac{z-1}{z}} = (1-S)^{\frac{1}{z}} - 1 + p$$

which has nontrivial solution $S > 0$ at $p > p_c = 1/(z-1)$.

Let us find $\langle s_k s_l \rangle$. We have

$$\frac{d^2 \tilde{G}(u(\mathbf{h}),\mathbf{h})}{dh_k dh_l} =$$

$$z(1-p)^z \left[\int_0^u dv \partial_k \partial_l A(v,\mathbf{h}) + p(1-p)^{z-2}\frac{A(u,\mathbf{h})\partial_k A(u,\mathbf{h})\partial_l A(u,\mathbf{h})}{A(u,\mathbf{h}) - u\partial_u A(u,\mathbf{h})}\right]_{u=u(\mathbf{h})}$$

$$\langle s_k s_l \rangle = \delta(k,l)\langle s_k \rangle + p(1-p)^{2z-2}\frac{kl}{z}\binom{z}{k}\binom{z}{l}u_0^{k+l-2}\frac{1+u_0}{1-(z-2)u_0} \qquad (17)$$

The last term here diverges at $p_c$ at all $k, l$ same as $\langle s^2 \rangle = \sum_{k=1}^{z}\sum_{l=1}^{z} \langle s_k s_l \rangle$.

In similar manner we can find other averages of $\prod_{k=1}^{z} s_k^{m_k}$ over $\nu_\mathbf{s}$. Note that such averages are taken over *all* finite clusters so, for example, $\langle s_k \rangle$ gives us the fraction of all graph's sites which belong to the finite clusters and have the degree $k$. However, it can be useful to consider instead the averages over degree distributions in the ensemble of finite clusters with fixed size $s = \sum_{m=1}^{z} s_m$, that is, over

$$\nu_{s,\mathbf{s}} = \nu_\mathbf{s} \delta\left(s, \sum_{m=1}^{z} s_m\right) \tag{18}$$

It appears that $\nu_{s,\mathbf{s}}$ can be obtained explicitly from (11, 13) for arbitrary $z$. To do this we change variables in the above equations

$$h_k = h'_k + \ln x, \quad \sum_{k=1}^{z} h'_k = 0, \quad x = \exp\left(z^{-1} \sum_{k=1}^{z} h_k\right),$$

In new variables the functions $A(u,\mathbf{h})$ and $G(u,\mathbf{h})$ do not change their functional form (11), (12) acquiring factor $x$ only

$$A(u,\mathbf{h}) = xA(u,\mathbf{h}'), \quad G(u,\mathbf{h}) = xG(u,\mathbf{h}')$$

Equation for $u(x,\mathbf{h}')$ becomes

$$u = p(1-p)^{z-2} xA(u,\mathbf{h}') \tag{19}$$

Consider the expansion of $G(\mathbf{h})$ in powers of $x$

$$\tilde{G}(\mathbf{h}) = \tilde{G}[u(\mathbf{h}),\mathbf{h}] = x\tilde{G}[u(x,\mathbf{h}'),\mathbf{h}'] = \sum_{s=1}^{\infty} \tilde{G}_s(\mathbf{h}') x^s.$$

According to (2) $\tilde{G}_s(\mathbf{h}')$ is generating function for the fixed-size degree distribution (18), that is

$$\tilde{G}_s(\mathbf{h}') = \sum_\mathbf{s} \nu_{s,\mathbf{s}} e^{\mathbf{sh}'} \tag{20}$$

As $\tilde{G}_s(\mathbf{h}')$ is the expansion coefficient of implicitly defined function $\tilde{G}[u(x,\mathbf{h}'),\mathbf{h}']$ and $u(x,\mathbf{h}')$ is analytic function of $x$ such that $u(0,\mathbf{h}') = 0$, $\left.\frac{\partial u(x,\mathbf{h}')}{\partial x}\right|_{x=0} \neq 0$, cf. Eq. (19), we can apply Lagrange inversion formula to express $\tilde{G}_s(\mathbf{h}')$ for $s > 1$ as [23]

$$\tilde{G}_s(\mathbf{h'}) = \frac{\left[p(1-p)^{z-2}\right]^{s-1}}{s-1} \oint \frac{du}{2\pi i} \frac{A(u,\mathbf{h'})^{s-1}}{u^{s-1}} \frac{\partial \tilde{G}(u,\mathbf{h'})}{\partial u} =$$
$$= \frac{zp^{s-1}(1-p)^t}{s(s-1)} \oint \frac{du}{2\pi i} \frac{A(u,\mathbf{h'})^s}{u^{s-1}}$$
(21)

Here $t = (z-2)s + 2$ is the empty edges' perimeter of s-site cluster [17].

The expansion of $A(u,\mathbf{h'})^s$ in powers of $u$ reads

$$A(u,\mathbf{h'})^s = s! \sum_{\mathbf{s}} \delta\left(\sum_{m=1}^{z} s_m, s\right) e^{\mathbf{h's}} \prod_{k=1}^{z} \binom{z-1}{k-1}^{s_k} u^{(k-1)s_k}$$

so the integration of $u^{1-s} A(u,\mathbf{h'})^s$ in (21) gives just the Kronecker delta

$$\delta\left(\sum_{m=2}^{z}(m-1)s_m + 2, s\right) = \delta\left(\sum_{m=1}^{z} ms_m + 2, 2s\right)$$

and according to (20) we finally get for s > 1

$$v_{s,\mathbf{s}} = p^{s-1}(1-p)^t D_{s,\mathbf{s}},$$
(22)

$$D_{s,\mathbf{s}} = z(s-2)! \delta\left(\sum_{m=1}^{z} ms_m + 2, 2s\right) \delta\left(\sum_{m=1}^{z} s_m, s\right) \prod_{k=1}^{z} \frac{1}{s_k!} \binom{z-1}{k-1}^{s_k}$$
(23)

Factor $p^{s-1}(1-p)^t$ gives the probability to find s-site cluster as each such cluster has $b = s-1$ bonds and $t$ empty edges touching it [17]. Delta symbol $\delta\left(\sum_{m=1}^{z} ms_m + 2, 2s\right)$ preserves the relation $b = s-1$ as $\sum_{m=1}^{z} ms_m = 2b$. Thus $D_{s,\mathbf{s}}$ is the number (per site) of s-site clusters with prescribed degree distribution **s**. Otherwise, it is the number of such clusters containing some given site divided by $s$. Note that $v_\mathbf{s} = \sum_{s=1}^{\infty} v_{s,\mathbf{s}}$ but to deal with $v_\mathbf{s}$ is not always convenient, it can be simpler to find the averages with fixed $s$ and then sum over $s$ if needed.

It is easy to show that Eqs. (22, 23) reproduce the known result for the clusters' size distribution for bond percolation on Bethe lattice [17]

$$v_s = \sum_{\mathbf{s}} v_{s,\mathbf{s}} = p^{s-1}(1-p)^t \frac{z}{st} \binom{s(z-1)}{s-1}$$

The summation here is easily performed using integral representation of Kronecker deltas. Similarly, we can find the average fraction of the degree $k$ sites in s-site cluster

$$\bar{w}_{s,k} = \sum_{\mathbf{s}} \frac{s_k}{s} \frac{v_{s,\mathbf{s}}}{v_s} = \binom{s-2}{k-1}\binom{t}{z-k}\binom{s(z-1)}{z-1}^{-1} \qquad (24)$$

This parameter diminishes monotonously with the growth of *k*. If $s < z+1$ then $\bar{w}_{s,k} = 0$ for $k > s-1$.

In the limit $s \to \infty$, $s_k / s \to w_k \neq 0$, $v_{s,s\mathbf{w}}$ always falls exponentially

$$v_{s,s\mathbf{w}} \approx \frac{z}{p}(1-p)^2 \left[ (2\pi)^{z-1} s^{z+3} \prod_{k=1}^{z} w_k \right]^{-\frac{1}{2}} \exp - s \sum_{k=1}^{z} w_k \ln \left[ \binom{z-1}{k-1}^{-1} \frac{w_k}{p(1-p)^{z-2}} \right]$$

except for the case of special $w_k^*$ which maximize $v_{s,s\mathbf{w}}$ with due constraints

$$w_k^* = \binom{z-1}{k-1}\frac{(z-2)^{z-k}}{(z-1)^{z-1}}, \qquad \sum_{k=1}^{z} w_k^* = 1, \qquad \sum_{k=1}^{z} k w_k^* = 2$$

In this case

$$v_{s,s\mathbf{w}^*} \approx zp(1-p)^2 \left[ (2\pi)^{z-1} s^{z+3} \prod_{k=1}^{z} w_k^* \right]^{-\frac{1}{2}} \left[ \frac{p(1-p)^{z-2}}{p_c(1-p_c)^{z-2}} \right]^s,$$

Thus, at $p_c$ $v_{s,s\mathbf{w}^*}$ exhibits only power law decay so large clusters with the degree distribution $w_k^*$ outnumber others at criticality. It seems natural to suggest that right above $p_c$ percolation cluster also have this degree distribution. Note that in the limit $s \to \infty$ the average $\bar{w}_{s,k}$ coincides with this critical $w_k^*$, cf. Eq. (24).

## IV. PERCOLATION ON COMPLETE GRAPH

Complete graph consists of *N* sites each pair of which is connected by an edge. The ensemble of bond percolation patterns on it with due weights is known as Erdos-Renyi (ER) random graph [7, 21]. Modified Potts model on complete graph is a mean-field model. In Hamiltonian (6) one should put $p = c/N$ to get meaningful thermodynamic limit. Then usual mean-field treatment gives

$$G(u,h_0,\mathbf{h}) = \tilde{G}(u,\mathbf{h}) + e^{-c+h_0}$$

$$\tilde{G}(u,\mathbf{h}) = \frac{e^{-c}}{2} \sum_{k=1}^{\infty} \frac{2-k}{k!} u^k e^{h_k} = \frac{e^{-c}}{2}\left[ 2\int_0^u dv A(v,\mathbf{h}) - uA(u,\mathbf{h}) \right]$$

where

$$A(u,\mathbf{h}) = \sum_{k=0}^{\infty} \frac{u^k e^{h_{k+1}}}{k!}$$

and $u = u(\mathbf{h})$ is defined through the equation

$$u = ce^{-c} A(u, \mathbf{h})$$

Again, differentiating $\tilde{G}[u(\mathbf{h}), \mathbf{h}]$ with respect to $\mathbf{h}$ we get the averages of powers of $s_k$ as in Eq. (14). We have

$$\langle s_k \rangle = e^{-c} \frac{u_0^k}{k!}, \quad u_0 = ce^{u_0 - c} \quad (25)$$

Introducing $S$ as the fraction of graph occupied by the giant component we get

$$u_0 = c(1-S), \quad 1-S = e^{-cS} \quad (26)$$

Eqs. (25, 26) are the standard relations for ER graph [7, 18, 21]. In particular, the nontrivial solution to Eq. (26) $S > 0$ appears at $c > 1$.

Derivation similar to that used to obtain Eq. (17) gives

$$\langle s_k s_l \rangle = \delta(k,l)\langle s_k \rangle + \frac{ce^{-2c} u_0^{k+l-2}}{(k-1)!(l-1)!(1-u_0)}$$

Also the described above procedure gives the clusters' size-degree distribution

$$v_{s,\mathbf{s}} = c^{s-1} e^{-cs} (s-2)! \delta\left(\sum_{m=1}^{\infty} m s_m + 2, 2s\right) \delta\left(\sum_{m=1}^{\infty} s_m, s\right) \prod_{k=1}^{\infty} \frac{1}{s_k! [(k-1)!]^{s_k}} \quad (27)$$

Here $s_k$ with arbitrary $k$ are present but actually constraints imposed by Kronecker deltas allow $s_k \neq 0$ for $1 \leq k \leq s-1$ only. Moreover, they restrict the $s_k$ values

$$s_k < \left[2\frac{s-1}{k}\right], \quad 1 \leq k, \quad s_1 < s$$

Here square brackets denote the integer part of a number.

With Eq. (27), one can reproduce the known result for clusters' size distribution of ER graph [18, 21]

$$v_s = \sum_{\mathbf{s}} v_{s,\mathbf{s}} = c^{s-1} e^{-sc} \frac{s^{s-2}}{s!}$$

Average fraction of degree $k$ sites in $s$-site clusters is

$$\bar{w}_{s,k} = \frac{(s-1)^{s-1-k}}{s^{s-2}} \binom{s-2}{k-1}, \quad 1 \leq k \leq s-1$$

For $s \to \infty$ $\bar{w}_{s,k} \to w_k^* = \frac{1}{e(k-1)!}$ which means that large clusters form the ensemble of the equilibrium random connected trees [7].

Note that all these results can be obtained from those for Bethe lattice in the limit $z \to \infty$, $zp \to c$, showing its equivalence to complete graph in the thermodynamic limit.

## V. DISCUSSION AND CONCLUSIONS

The realization of present method can give some insight into the structure of finite clusters in classical bond percolation as degree distribution also gives the average number of bonds $2b = \sum_{k=1}^{z} k \langle s_k \rangle$, the number of perimeter sites $s(1 - \bar{w}_{s,z})$ and fractal dimension $1/d = 1 + \lim_{s \to \infty} \ln(1 - \bar{w}_{s,z}) / \ln s$ along with some notion of the cluster's *k*-core decomposition [7, 21] as the *k*-core fraction $n_k \leq \sum_{m=k}^{z} \bar{w}_{s,m}$. However, we cannot say how the sets of *k*-degree sites are connected to each other and inside themselves, e. g., how many *k*-cliques they contain [21]. Probably, the combination of the present approach and those of Ref. [24] could help to get such information.

The other problem is how to apply the present method to many models of correlated percolation [1, 2, 7, 21]. This seems rather obvious for the statistics of like-sign clusters in Ising models [25] but not for the cases of, say, bootstrap or explosive percolation. The main obstacle is again the absence of information on the connectivity of *k*-degree subsets.

Yet, present approach gives the new data on clusters in classical bond percolation on the simple graphs and provides the tool for the approximate analytical or numeric description of clusters' degree distributions on more realistic lattices and graphs. It may have some perspectives in the studies of classical percolation as well as the variety of correlated percolation models.

## ACKNOWLEDGEMENT

The author acknowledges support by the Ministry of Education and Science of the Russian Federation (state assignment grant No. 3.5710.2017/BCh).